\documentclass{ws-procs9x6}
\begin{document}

\title{Impact on Astrophysics and Elementary Particle Physics\\
of recent and future Solar Neutrino data}

\author{V. Antonelli$^*$ and L. Miramonti}

\address{Physics Department, Milano University and I.N.F.N. Sezione di Milano,\\
Via Celoria 16, 20133-Milano, Italy\\
$^*$E-mail: vito.antonelli@mi.infn.it}

\begin{abstract}
The study of neutrinos is fundamental to connect astrophysics and elementary particle physics. 
In this last decade solar neutrino experiments and KamLAND 
confirmed the LMA solution and further 
clarified the 
oscillation pattern. Borexino attacked also the study of the low energy neutrino spectrum. 
However, important points still 
need clarification, like the apparent anomaly in the vacuum to matter 
transition region. Besides, a more detailed study of the low energy components of the pp 
cycle, combined with a measurement of CNO fluxes, is compulsory, also to 
discriminate between the low and the high Z versions of the Solar Standard 
Models and solve the metallicity problem. 
We discuss the main recent advancements 
and the possibilities of studying these open problems with 
Borexino, SNO+ and the future experiments, like the next generation of scintillators.
\end{abstract}


\bodymatter\bigskip

Astrophysics and elementary particle physics are strictly connected since ever. 
Many discoveries of the intimate structure of matter had an important impact on cosmology and astrophysics and at the same time the study of stars, galaxies and of the large scale structure of the universe helped in the comprehension of the fundamental interactions and of elementary particles classification and properties. The cosmic rays, for instance, were essential for the knowledge of elementary particles and also of the processes taking place in the space. 
In our days we can recall the double role of LHC studies. The Higgs discovery completed the consistency proof of Standard Model and of the mass generation mechanism and the search for new high energy particles could give the first proof of supersymmetry  or indicate the need to modify these theories. 
At the same time the possibility of testing at LHC higher and higher energies offers a unique opportunity to approach conditions closer to the ones of the first stages after the Big Bang and get important information about the universe evolution. 
Another example is given by the studies of the Cosmic Microwave Background. 
The data collected by COBE~\cite{COBE}, BOOMERanG~\cite{Boomerang} and 
WMAP~\cite{WMAP} and the recent data from Planck~\cite{Planck1, Planck2, 
Planck3, PlanckBersanelli} had a great impact on cosmological theories about the universe structure and evolution, but, as a fundamental by product, they also contributed to elementary particle physics (putting upper limits on the absolute neutrino mass competitive with the ones recovered by direct searches).
Other experiments, like Auger~\cite{Auger013, Auger010}, studying very high energy cosmic rays and neutrinos,
are  expected to give us a better insight in the mechanisms of cosmic production and acceleration of particles, but they also offer a unique possibility of testing the neutrino oscillation mechanism at
energies higher than the ones reachable with terrestrial neutrino sources.
The study of neutrino
properties is a typical example of interplay between elementary particle physics and astrophysics. The proof that neutrinos are massive and oscillation particles, the first clear experimental indication of the need to go beyond the ``minimal version'' of the Standard Model of electroweak interactions, has been obtained mainly studying neutrinos from the atmosphere, the Sun and Supernova explosions; meanwhile 
neutrino studies made possible important steps forward in the knowledge of all of these cosmological objects. 

Solar neutrino 
studies started in the late sixties (thanks to Bahcall's idea of studying the interior of the Sun by detecting the neutrinos it emits) and with the pioneering Homestake experiment~\cite{Homestake} . The results gave rise to the ``Solar Neutrino Problem'' (SNP), which found a final solution only four decades later, when the SNO collaboration\cite{SNONC} and the reactor experiment KamLAND~\cite{Kamlandprimo}, together with previous 
experiments, proved clearly that neutrinos are massive and oscillating particles and confirmed the validity of the Solar Standard Model 
(SSM)~\cite{GS98,AGSS09} . 
For a recent review on these subjects 
we refer the reader to~\cite{Noi} .

In this last decade the main attention of 
neutrino
community moved to 
artificial 
sources (from reactors and accelerators) and/or 
appearance signals. However, important steps forward have been done also in solar neutrino physics, with many data and analyses collected by ``historical experiments'' and above all with the Borexino 
advent~\cite{Borexino} . 
The original Bahcall's dream is becoming true, but important questions are still waiting for an answer, of great relevance for elementary particle physics and astrophysics. 
%
A more detailed analysis of transition region between the vacuum dominated and the matter enhanced part of solar 
neutrino spectrum 
could fully confirm the LMA description or the presence of anomalies and 
a deep investigation of the lower energy part of pp spectrum together with the CNO bicycle
could significantly contribute to solve the ``metallicity problem'', discriminating between high-Z~\cite{GS98} and low-Z~\cite{AGSS09} versions of SSM.
       
Recently the SNO collaboration performed a combined analysis of all the data obtained 
by the experiment~\cite{SNOrecente} : SNO-I and SNO-II (the '01-'03 salt phase), already combined by the LETA analysis\cite{SNOLETA}, and SNO-III\cite{SNOelio} ('03-'06 data with $^3$He filled proportional chambers). 
The value of $^8$B $\nu$ flux, 
$\phi_{^8\rm{B}} =\left(5.25 \pm 0.16 (stat)^{+0.11}_{-0.13} (syst) \right)
\times 10^6$ cm$^{-2}$ s$^{-1}$ was in agreement with previous results, but with a lower indetermination. 
Unfortunately it fell in the middle between the predictions of the 
high-Z~\cite{GS98} ($\phi_{^8\rm{B}}=(5.88\pm0.65)\times10^6$) 
and low-Z~\cite{AGSS09} 
($\phi_{^8\rm{B}}=(4.85\pm0.58)\times10^6$) models; therefore it cannot discriminate between them. 
The Day-Night (D-N) asymmetry was compatible with zero and there was no anomalous seasonal variation. 
Also SuperKamiokande (SK) went on taking data, with SK-II\cite{SKII} ('02-'05), SK-III\cite{SKIII} ('06-'08) and SK-IV, started in september 2008. The first SK-IV results, presented  at this conference,
\cite{SKComo} 
seem to show the presence of the expected spectrum distortion below 6 MeV (even if it is only at 1$\sigma$ at the moment) and to give a significant (2.7 $\sigma$) indication of a D-N asymmetry different from zero (compatible with a $\nu_e$ regeneration in Earth).

Meanwhile KamLAND (KL) collected from 2004 to '07 data with an enlarged fiducial volume and values of the systematic uncertainties and the background reduced with respect to the ones of 
the '02-'04 campaign. These results\cite{KL2007} were in general agreement with the solar $\nu$ data, with a slight tension of KL toward higher values of $\Delta m_{21}^2$ and tan$^2 \theta_{12}$\cite{Petcov}, which 
was anyhow reduced in a three flavours analysis.
The global analysis performed by SNO collaboration gave the following output for the mixing and mass parameters~\cite{SNOrecente}:
$\Delta m_{21}^2=\left(7.41^{+0.21}_{-0.19}\right)\times10^{-5}\rm{eV}^2; \, \rm{tan}^2 \theta_{12}=0.446^{+0.030}_{-0.029}; \, \rm{sin}^2 \theta_{13}=(2.5^{+1.8}_{-1.5})\times10^{-2}$, 
significantly confirming that $\theta_{13} \neq 0$, as recently proved by reactor \cite{theta13-reactor} and accelerator\cite{theta13-accelerator} experiments and already indicated by global phenomenological analyses, like\cite{LISI}. 
A recent global analysis performed by the KL collaboration,
including the     
$\theta_{13}$ constraints from these experiments, found similar values for the two other parameters. 

SNO, SK and KamLAND could investigate only the higher energy part of pp spectrum (more or less above 4 MeV) and up to five years ago the low energy part, representing the main component of the spectrum, had been studied only by the radiochemical experiments, that can measure only the integrated spectrum.
Borexino was the first real time experiment to explore (starting from '07) the sub-MeV region and to isolate the monochromatic Berillium line. 
After the 2009 calibration campaign, 
the $^7$Be $\nu$ flux was accurately measured~\cite{BorexinoBe7},  
$\phi_{^7\rm{Be}}=(3.10\pm0.15)\times10^9$ cm$^{-2}$ s$^{-1}$, excluding the no oscillation hypotheses 
at more than $5 \sigma$. 
Borexino measured also the $^8$B flux\cite{BorexinoB8} with a threshold of about $3$ MeV for the recoil 
electron kinetic energy and its result was compatible with the ones of previous solar neutrino
experiments. 
The absence of D-N asymmetry
confirmed the LMA solution of the SNP.
The direct measurement of $^7$Be flux was very important, but it still doesn't 
allow discrimination between high-Z and low-Z models, which predict values for this flux 
respectively higher and lower than Borexino result, but in both cases compatible with it, 
as it was in the case of $^8$B $\nu$ flux. 
As discussed in the first paper of~\cite{Borexino}, it would be fundamental to improve the discriminating power of solar neutrino experiments. An analysis of the $^7$Be and $^8$B components will probably not be sufficient; a combined study of these fluxes (mainly $^8$B) and of CNO neutrinos could, instead, 
try to determine which is the right version of Solar Models and to solve the ``metallicity problem''.    

In the SSM the pep neutrino
flux is strongly constrained (uncertainty around $1.2 \%$) by the solar luminosity and the link to pp component; therefore this flux is the better probe (after pp) 
to test the models. The study of CNO neutrinos can contribute significantly to the determination of the solar core metallicity and moreover this cycle is considered fundamental to fuel 
stars more massive than the Sun. 
Nevertheless, no direct detection of pep and CNO 
neutrinos was available until 2011. To face the problem of background, mainly from $\beta^+$ 
emitter $^{11}$C, Borexino developed the special TFC (Three Folder Coincidence) technique, making possible the measurement (2007-'10) of   
the pep 
flux 
$\phi_{\rm{pep}}=(1.6\pm0.3)\times10^8\rm{cm}^{-2}\rm{s}^{-1}$ and of the upper 
limit 
for CNO neutrinos, $\phi_{\rm{CNO}}<7.7\times10^8\rm{cm}^{-2}\rm{s}^{-1}$ 
(at 95$\%$ C.L.))~\cite{BorexpepCNO}.
%

An improvement of pep neutrinos
meaurement (and possibly a direct determination of the 
pp ones) would be a stringent consistency test of SSMs.  
A full comprehension of the oscillation would also require further investigation of the transition between the vacuum and matter enhanced regions of the energy spectrum, corresponding to the lower energy part of $^8$B flux. 
SNO and Borexino seem to indicate a partial absence of the upturn predicted at low energies by the LMA solution, but this hint is not confirmed by the recent SK data (even if their statistical significance is lower)\cite{SKComo}.

Water Cerenkov  
can detect only the medium and higher energy part of the spectrum; radiochemical experiments, 
instead, can measure only integrated rates. The most important contributions 
should come by scintillators, ideal for low energy solar neutrino spectroscopy, thanks to  their large masses and high photon yield and radiopurity levels. Results are expected by Borexino 
and SNO+, a multipurpose liquid scintillator experiment at SNO-LAB, which is nearing its final phase of construction~\cite{SNO+} . 
SNO+ will look for neutrinoless double $\beta$ decay ($0 \nu 2 \beta$), but it could also measure geoneutrinos, neutrinos from reactors and Supernovae and from the Sun (the lower energy part of pp spectrum and the CNO cycle).   
The larger mass and a position almost twice deeper than the Borexino one should guarantee a better signal to background ratio for pep detection, 
with the possibility of reaching a 
$5 \%$ accuracy. 

Future solar neutrino experiments, like the ones looking for 
$0 \nu 2 \beta$ decays and dark matter. 
should be able to detect very low signal at low energies and, therefore, to reduce as much as possible the background.
The solution could be offered by multipurpose detectors with large masses and very low radiopurity levels, 
like the next generation of scintillators, varying from the organic scintillators with new technological devices to the ones using new materials (noble gases).
An example of the first kind is LENA~\cite{LENA}, 
a $50 \, \rm{ktons}$ liquid scintillator that will be hosted in the Physalmi underground laboratories to study the $\nu-e^{-}$ elastic scattering and the charged current 
($\nu_e + ^{13}C \to ^{13}N + e^-$).
It could study in detail the $^7$Be signal (including the signal modulation) and the low 
energy $^8$B $\nu$ and attack the problem of pep and CNO detection. Future liquid scintillators 
with noble gases (Xe, Ar, Ne) should make possible the realization of large homogeneous detectors, quite easy to purify with an high scintillation yield and no autoabsorption.
%
This is the case of CLEAN/DEAP~\cite{CLEAN} 
and XMASS\cite{XMASS} detectors, whose prototypes have been tested, with a scalable tecnology, repectively in SNO-LAB and in Kamioka. In addition to $^7$Be $\nu$, they aim to measure pp flux with a $1 \%$ statistical uncertainty and CNO flux with a $10-15 \%$ accuracy. 
%
Interesting results could be obtained also by the EU-US working group DARWIN, that is focusing on a multiton liquid Ar (or Xe) detector for dark matter searches; the low energy $\nu-e^-$ elastic scattering would be one of the main background for this detector. 
A different strategy will be adopted by experiments planning to study the lower pp cycle components by means of inverse $\beta$ decay on metals like Indium, in the case of LENS~\cite{LENS} (the forerunner of this class) and IPNOS~\cite{IPNOS} , or Molybdenum (in the case of MOON~\cite{MOON}). 


\end{document}